\documentclass[reprint,twocolumn,showpacs,showkeys,superscriptaddress,preprintnumbers,amsmath,amssymb,floatfix,aps,pra]{revtex4-1}

\usepackage{graphicx}
\usepackage{bm}

\usepackage{placeins}
\usepackage{graphics}
\usepackage{color}
\usepackage{hyperref}
\usepackage{multirow}
\usepackage{pdfpages}

\makeatletter
\AtBeginDocument{\let\LS@rot\@undefined}
\makeatother

\begin{document}


\title{Strain-tunable orbital, spin-orbit, and optical properties of monolayer transition-metal dichalcogenides}

\author{Klaus Zollner}
\email{klaus.zollner@physik.uni-regensburg.de}
\affiliation{Institute for Theoretical Physics, University of Regensburg, 93040 Regensburg, Germany}

\author{Paulo E. Faria~Junior}
\affiliation{Institute for Theoretical Physics, University of Regensburg, 93040 Regensburg, Germany}

\author{Jaroslav Fabian}
\affiliation{Institute for Theoretical Physics, University of Regensburg, 93040 Regensburg, Germany}

\date{\today}

\begin{abstract}
When considering transition-metal dichalcogenides (TMDCs) in van der Waals (vdW) heterostructures 
for periodic \textit{ab-initio} calculations, usually, lattice mismatch is present, and the TMDC 
needs to be strained. In this study we provide a systematic assessment of biaxial strain effects 
on the orbital, spin-orbit, and optical  properties of the monolayer TMDCs using \textit{ab-initio} calculations. 
We complement our analysis with a minimal tight-binding Hamiltonian that captures the low-energy 
bands of the TMDCs around K and K' valleys. We find characteristic trends of the orbital and 
spin-orbit parameters as a function of the biaxial strain. Specifically, the orbital gap decreases 
linearly, while the valence (conduction) band spin splitting increases (decreases) nonlinearly in 
magnitude when the lattice constant increases. Furthermore, employing the Bethe-Salpeter equation  
and the extracted parameters, we show the evolution of several exciton peaks, with 
biaxial strain, on different dielectric surroundings, which are particularly useful for 
interpreting experiments studying strain-tunable optical spectra of TMDCs.
\end{abstract}

\maketitle

\section{Introduction}
\label{sec:Introduction}

A vastly evolving field of condensed matter physics is that of two-dimensional (2D) van der Waals (vdW) materials and their hybrids. 
The available material repertoire covers semiconductors \cite{Kormanyos2014:2DM, Liu2015:CSR, Tonndorf2013:OE,Tongay2012:NL,Eda2011:NL} (MoS$_2$, WSe$_2$), ferromagnets \cite{Li2014:JMCC, Carteaux1995:JP, Gong2017:Nat, Siberchicot1996:JPC, Lin2017:PRB, Wang2018:NN, Liu2016:PCCP, Zhang2015:JMCC, McGuire2015:CoM, Webster2018:PCCP, Huang2017:Nat, Jiang2018:NL, Soriano2018:arxiv,Huang2018:NN, Jiang2018:NN, Wu2019:NC} (CrI$_3$, CrGeTe$_3$), superconductors \cite{Yoshida2016:APL, Noat2015.PRB, Zhu2016:NC} (NbSe$_2$), and topological insulators \cite{Xu2018:NP} (WTe$_2$), which offer unforeseen potential for 
electronics and spintronics \cite{Fabian2007:APS, Zutic2004:RMP}. 
For example, monolayer transition-metal dichalcogenides (TMDCs) are direct band gap semiconductors with remarkable physical properties \cite{Mak2010:PRL,Chernikov2014:PRL, Kormanyos2014:2DM, Liu2015:CSR, Tonndorf2013:OE,Tongay2012:NL,Eda2011:NL, Gibertine2014:PRB,Wang2018:RMP}, 
specially in the realm of optoelectronics \cite{Wang2012:NN},  
optospintronics \cite{Gmitra2015:PRB,Luo2017:NL,Avsar2017:ACS}, and valleytronics \cite{Schaibley2016:NRM, Langer2018:Nat, Zhong2017:SA}. 
Currently, TMDCs, being stable in air, are a favorite platform for optical experiments including optical spin injection due to helicity-selective optical excitations \cite{Xiao2012:PRL}.

The ability to control and modify the electronic, spin, and optical 
properties of 2D materials is extremely valuable for investigating novel physical phenomena, as well as a potential knob for device applications. 
One possibility to do so in TMDCs is by deforming the crystal 
lattice via strain engineering \cite{He2013:NL,Conley2013:NL,Plechinger2015:2DMat,Ji2016:CPB,Schmidt2016:2DM,Lloyd2016:NL,Aslan2018:PRB,Aslan2018:NL,Frisenda2017:2DMA,Gant2019:MatToday,Tedeschi2019:AM, Iff2019:NL, Blundo2019:arXiv}. Recent experiments have shown that strain modulation 
is very effective and can lead to changes in the optical transition energies by hundreds of meV with just a few percent of applied strain \cite{Lloyd2016:NL, He2013:NL, Conley2013:NL, Aslan2018:PRB, Aslan2018:NL, Schmidt2016:2DM}. Even more interesting is that this strain modulation is completely reversible \cite{Lloyd2016:NL,Schmidt2016:2DM}.
As a general trend observed in the experimental studies, biaxial strain induces a significantly stronger modulation when compared to uniaxial strain, a fact also supported by {\it ab-initio} calculations \cite{Peelaers2012:PRB, Johari2012:ACS}.
Furthermore, by strain engineering it is possible to localize excitons in specific regions, which is a viable approach to obtain spatially and spectrally isolated quantum emitters based on 2D materials \cite{Castellanos-Gomez2013:NL,Kumar2015:NL,Branny2016:APL,Proscia2018:Optica,Iff2019:NL}.

Strain also plays an important role when TMDCs are stacked on or sandwiched by other 2D materials creating vdW heterostructures \cite{Novoselov2016:SC, Geim2013:Nat}. An example of interesting physics present in vdW heterostructures are the proximity effects \cite{Zutic2018:MT}. Typical examples involving TMDCs are: spin-orbit coupling (SOC) induced in graphene by TMDCs \cite{Gmitra2015:PRB,Gmitra2016:PRB} and proximity exchange induced in the TMDC due to magnetic substrates \cite{Zollner2019:PRB, Seyler2018:NL, Zhong2017:SA, Qi2015:PRB}. 

Strain effects are extremely important from a theoretical point of view: by creating vdW heterostructures that fulfill the periodic boundary conditions of first-principles calculations, it is often necessary 
to adjust the lattice parameters of the materials involved, therefore leading to strained crystals. Certainly, the strain --- which is biaxial in first-principles calculations --- will modify the electronic structure of the TMDC and therefore a systematic analysis of its behavior can provide 
valuable insight not only from an experimental point of view but also to aid in the design of novel heterostructures.

In this paper, we study the effect of biaxial strain on the orbital, spin-orbit, and optical properties of pristine monolayer TMDCs. We find that by tuning the lattice constant, the orbital band gap, and 
the spin splittings of the valence and conduction bands drastically change. 
Specifically, the orbital gap decreases linearly, while the valence (conduction) band spin splitting
increases (decreases) nonlinearly in magnitude, when the lattice constant increases. 
The observed behavior is universal for all studied TMDCs (MoS$_2$, MoSe$_2$, WS$_2$, WSe$_2$).
In addition, we show that spin splittings of the bands result from an interplay of the atomic SOC values of the transition-metal and chalcogen atoms. 
Finally, we analyze the direct-indirect transition energies and by employing the Bethe-Salpeter equation we calculate the optical absorption 
spectra of the biaxially strained TMDC monolayers. We show the evolution of several exciton peaks and 
their energy differences as a function of strain, assuming different dielectric surroundings. We also extracted the gauge factors --- the rates at which the exciton peak energies shift due to strain --- which are relevant for comparison to experiments.

\section{Model Hamiltonian}
\label{sec:Hamiltonian}

\begin{figure}[htb]
 \includegraphics[width=.99\columnwidth]{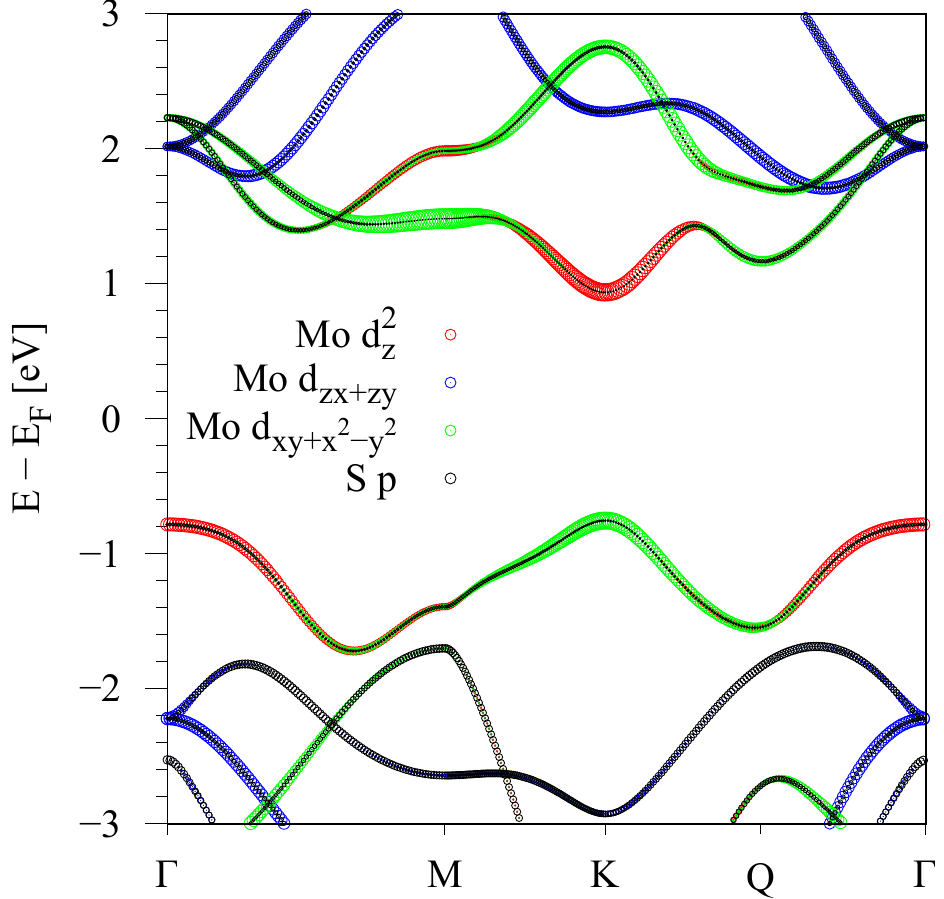}
 \caption{(Color online) Calculated orbital decomposed 
 band structure of MoS$_2$, as a representative example of a TMDC. 
 SOC is not included and the different colors correspond to different orbitals or atoms. 
 }\label{Fig:bands_TMDC_character}
\end{figure}

In the manuscript, we deal with TMDC monolayers. 
Therefore we need a Hamiltonian that describes
the low energy bands of bare TMDCs around the K and K' valleys, including spin-valley locking. 
In Fig.~\ref{Fig:bands_TMDC_character} we show the orbital
decomposed band structure of MoS$_2$ without inclusion of SOC, 
as a representative example of a 
TMDC with general structure MX$_2$ (M for the transition metal atom, X for the chalcogen atom).
The wave functions we use for the Hamiltonian are 
$|\Psi_{\textrm{CB}}\rangle = |d_{z^2}\rangle$ and 
$|\Psi_{\textrm{VB}}^{\tau}\rangle = \frac{1}{\sqrt{2}}(|d_{x^2-y^2}\rangle+\textrm{i}\tau |d_{xy}\rangle)$, 
corresponding to the conduction band (CB) and the valence band (VB) at K and K', since the band edges 
are formed by different $d$-orbitals from the transition metal, see Fig.~\ref{Fig:bands_TMDC_character}, in agreement 
with literature \cite{Kormanyos2014:2DM}.
The model Hamiltonian to describe 
the band structure (including SOC) of the TMDC close to K ($\tau = 1$) and K' ($\tau = -1$) is
\begin{flalign}
\label{Eq:Hamiltonian}
&\mathcal{H} = \mathcal{H}_{0}+\mathcal{H}_{\Delta}+\mathcal{H}_{\textrm{soc}},\\
&\mathcal{H}_{0} = \hbar v_{\textrm{F}} s_0 \otimes (\tau\sigma_{x}k_{x}+\sigma_{y}k_{y}),\\
&\mathcal{H}_{\Delta} = \frac{\Delta}{2}s_{0}\otimes \sigma_{z},\\
&\mathcal{H}_{\textrm{soc}} = \tau s_{z} \otimes (\lambda_{\textrm{c}}\sigma_{+} + \lambda_{\textrm{v}} \sigma_{-}).
\end{flalign}
Here, $v_{\textrm{F}}$ is the Fermi velocity and
the Cartesian components $k_{x}$ and $k_{y}$ of the electron wave vector are measured from K~(K'). 
The pseudospin Pauli matrices are $\sigma_{\textrm{i}}$ acting on
the (CB,VB) subspace and spin Pauli matrices are $s_{\textrm{i}}$ 
acting on the ($\uparrow,\downarrow$) subspace,
with $\textrm{i} = \{0,x,y,z\}$. For shorter notation we introduce $ \sigma_{\pm} = \frac{1}{2}(\sigma_{0} \pm \sigma_{z})$.
TMDCs are semiconductors, and thus $\mathcal{H}_{\Delta}$ introduces a gap, represented by parameter $\Delta$, 
in the band structure such that $\mathcal{H}_{0}+\mathcal{H}_{\Delta}$ 
describes a gapped spectrum with spin-degenerate parabolic
CB and VB. In addition the bands are spin-split due to SOC which is captured by 
the term $\mathcal{H}_{\textrm{soc}}$ with the parameters $\lambda_{\textrm{c}}$ and $\lambda_{\textrm{v}}$ describing 
the spin splitting of the CB and VB.
The Hamiltonian $\mathcal{H}_{0}+\mathcal{H}_{\Delta}+\mathcal{H}_{\textrm{soc}}$ is already suitable to describe
the spectrum of bare TMDCs around the band edges at K and K'. 
The four basis states are $|\Psi_{\textrm{CB}}, \uparrow\rangle$, 
$|\Psi_{\textrm{VB}}^{\tau}, \uparrow\rangle$, $|\Psi_{\textrm{CB}}, \downarrow\rangle$, 
and $|\Psi_{\textrm{VB}}^{\tau}, \downarrow\rangle$. 
From now on, we consider only first-principles results, where SOC is included. 

\section{Geometry, Band Structure, and Fitted results}
\label{sec:Monolayer_TMDC}
To study proximity effects in TMDCs, one has to interface them with other materials, for example CrI$_3$ to get proximity exchange \cite{Zollner2019:PRB}. In these heterostructures, usually lattice mismatch between the constituents is present, and
we have to find a common unit cell for them,
to be applicable to periodic DFT calculations.
The usual approach is to create supercells of the individual materials, such that they can form a common unit cell, and strain is minimized.
Therefore, we introduce biaxial strain on the TMDC lattice, up to a reasonable limit, in heterostructure calculations. 
An important question is, whether the biaxial strain, will influence the \textit{intrinsic} properties, such as 
orbital gap and spin-orbit splittings, of the TMDC. 
Therefore, we calculate the band structures of the monolayer
TMDCs in a $1\times 1$ unit cell for different lattice constants, 
corresponding to biaxial strain with a maximum of $\pm 3\%$. 

The electronic structure calculations and structural relaxation of 
our geometries are performed with density functional theory (DFT)~\cite{Hohenberg1964:PRB}
using \textsc{Quantum Espresso}~\cite{Giannozzi2009:JPCM}.
Self-consistent calculations are performed with the $k$-point 
sampling of $30\times 30\times 1$ for bare TMDC monolayers. 
We use an energy cutoff for charge density of $560$~Ry, and
the kinetic energy cutoff for wavefunctions is $70$~Ry for the scalar relativistic pseudopotential 
with the projector augmented wave method \cite{Kresse1999:PRB} with the 
Perdew-Burke-Ernzerhof (PBE) exchange correlation functional \cite{Perdew1996:PRL}.
When SOC is included, the fully relativistic versions of the pseudopotentials are used. 
In order to simulate quasi-2D systems, a vacuum of at least $16$~\AA~is used 
to avoid interactions between periodic images in our slab geometries.
Structural relaxations of the monolayers, are performed with a quasi-Newton algorithm based on the 
trust radius procedure, until all components of all forces 
are reduced below $10^{-4}$~[Ry/$a_0$], where $a_0$ is the Bohr radius.

\begin{figure}[htb]
 \includegraphics[width=.99\columnwidth]{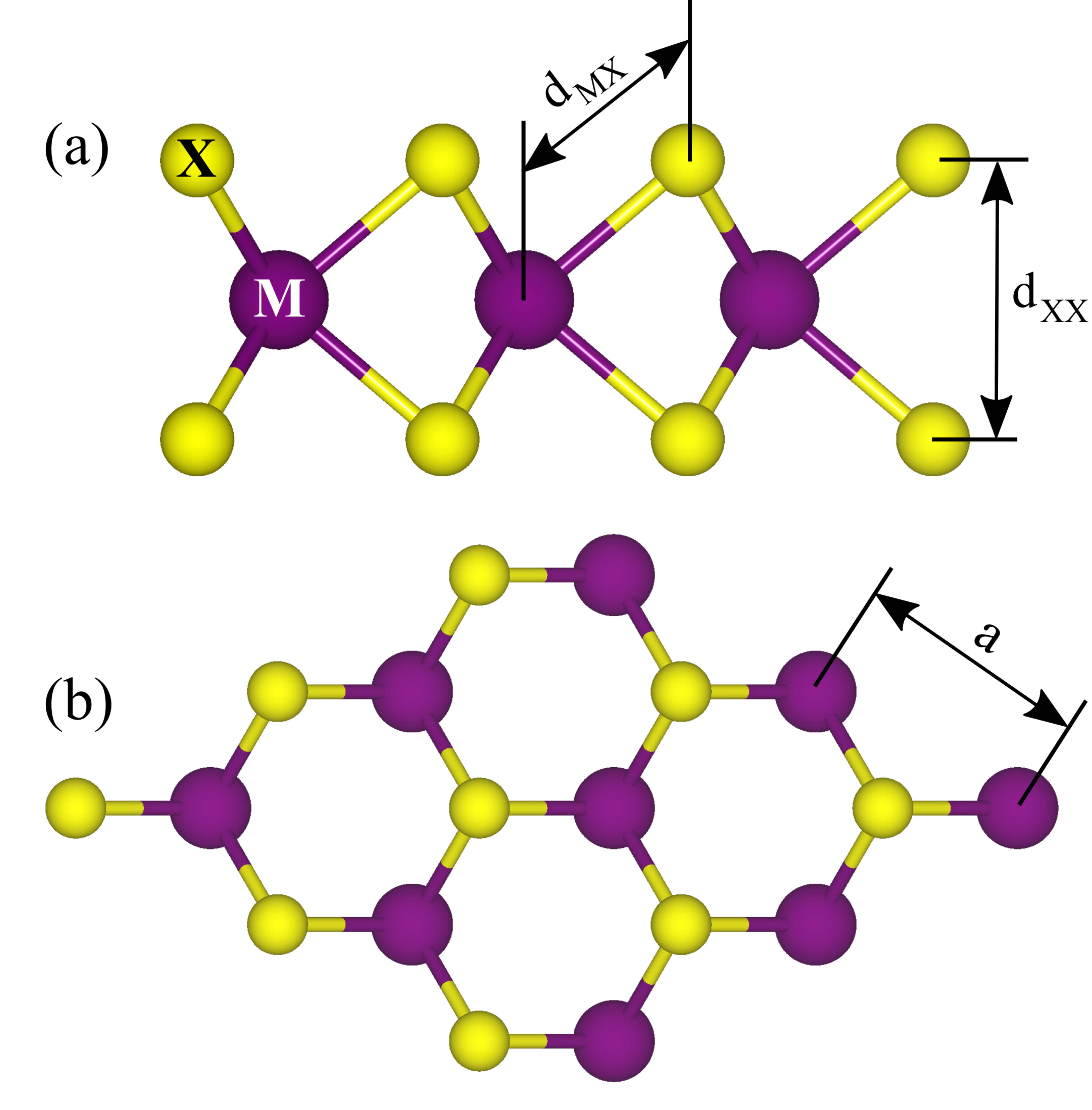}
 \caption{(Color online) Geometry of a TMDC monolayer with general structure MX$_2$, where
M is the transition metal (Mo, W) and X is the chalcogen atom (S, Se). 
(a) Side and (b) top view of the geometry, with labels for the lattice constant $a$, 
distance $d_{\textrm{MX}}$ (between the transition metal and the chalcogen atom),
 and $d_{\textrm{XX}}$ (between the two chalcogen atoms).
 }\label{Fig:structure_MX2}
\end{figure}

In Fig. \ref{Fig:structure_MX2} we show the geometry of a 
TMDC monolayer with general structure MX$_2$, where
M is the transition metal (Mo, W) and X is the chalcogen atom (S, Se). 
The distance between two chalcogen atoms is $d_{\textrm{XX}}$, 
the distance between the transition metal and the chalcogen atom is $d_{\textrm{MX}}$, 
and the distance between two transition metal atoms is the 
lattice constant $a$. We consider a series of lattice constants,
close to the experimental and theoretically predicted values 
of each TMDC, as summarized in Table~\ref{Tab:alat_TMDC}.

\begin{table}[htb]
\caption{Overview of the lattice parameters for all TMDCs, 
as well as fit parameters of the Hamiltonian $\mathcal{H}_{0}+\mathcal{H}_{\Delta}+\mathcal{H}_{\textrm{soc}}$.
The monolayer 
calculated lattice constant $a$ (calc.), distances $d_{\textrm{XX}}$, and $d_{\textrm{MX}}$, 
as defined in Fig~\ref{Fig:structure_MX2}. The orbital gap parameter $\Delta$, the Fermi velocity $v_{\textrm{F}}$ and 
the SOC parameters $\lambda_{\textrm{c}}$ and $\lambda_{\textrm{v}}$.
The experimental lattice constants $a$ (exp.) \cite{Wakabayashi1975:PRB, Schutte1987:JSSC, James1963:AC}
of the bulk systems are given for comparison.}
\label{Tab:alat_TMDC}
\begin{ruledtabular}
\begin{tabular}{lcccc}
 & MoS$_2$ & WS$_2$ & MoSe$_2$ & WSe$_2$\\
 \colrule
 $a$ (exp.) [\AA] & 3.15 & 3.153 & 3.288 & 3.282 \\
 $a$ (calc.) [\AA] & 3.185  & 3.18 & 3.319 & 3.319 \\
 $d_{\textrm{MX}}$ (calc.) [\AA] & 2.417 & 2.417 & 2.547 & 2.550 \\
 $d_{\textrm{XX}}$ (calc.) [\AA] & 3.138 & 3.145 & 3.357 & 3.364\\
 $\Delta$ [eV] &  1.687 & 1.812 & 1.461 &  1.525 \\
 $v_{\textrm{F}}$ [$10^{5} \frac{\textrm{m}}{\textrm{s}}$] & 5.338 & 6.735 & 4.597 & 5.948 \\
 $\lambda_{\textrm{c}}$ [meV]  & -1.41 & 15.72 & -10.45 & 19.86\\
 $\lambda_{\textrm{v}}$ [meV]  &  74.6 & 213.46 & 93.25 & 233.07\\
\end{tabular}
\end{ruledtabular}
\end{table}

\begin{figure}[htb]
 \includegraphics[width=.99\columnwidth]{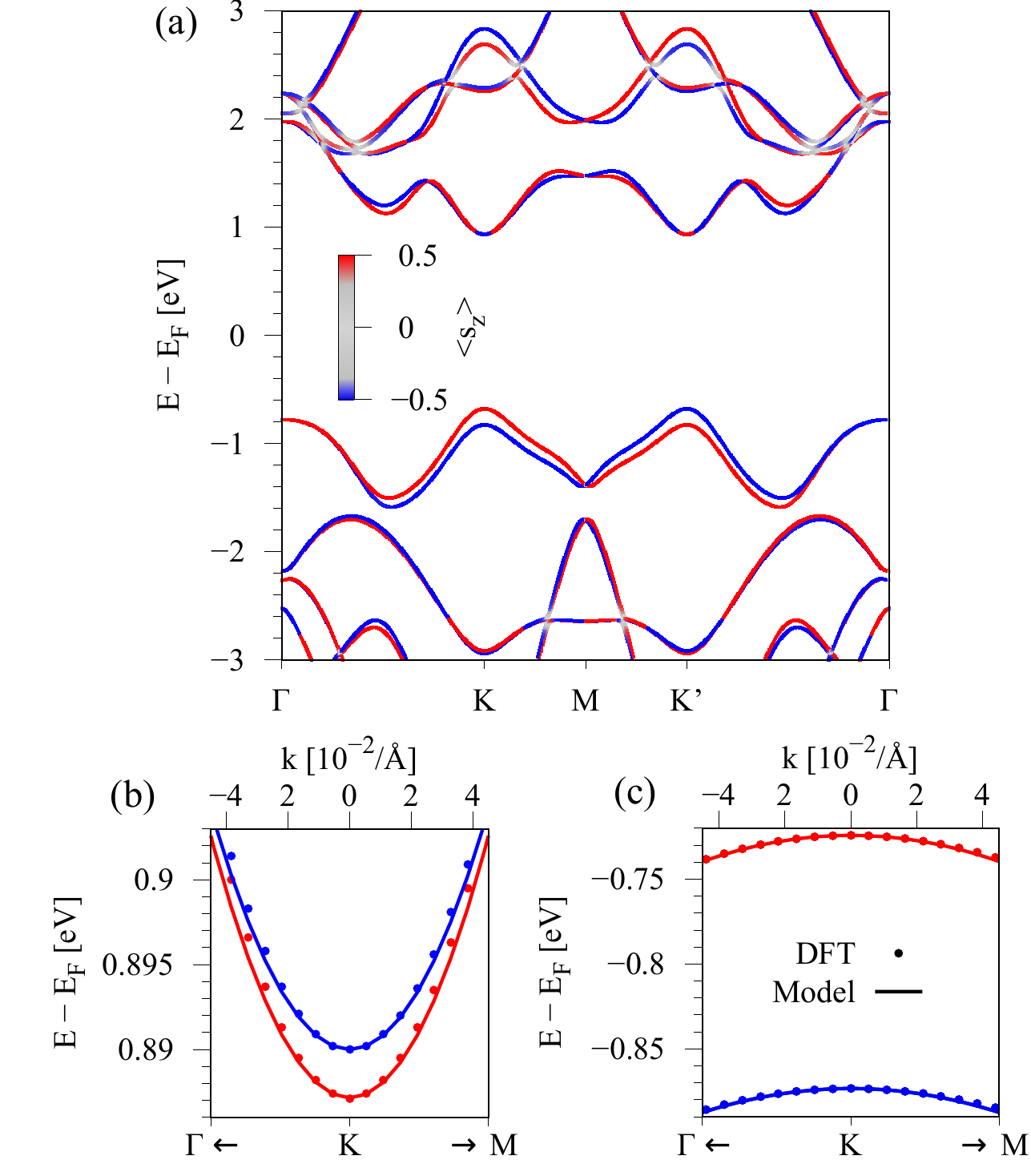}
 \caption{(Color online) (a) Calculated band structure of MoS$_2$ including SOC. 
 The color corresponds to the $s_z$ expectation value.  
 (b,c) Calculated low energy CB and VB around the K point (symbols) 
 with a fit to the model Hamiltonian (solid line).}
 \label{Fig:bands_TMDC_fit}
\end{figure}

\begin{figure*}[htb]
 \includegraphics[width=0.8\textwidth]{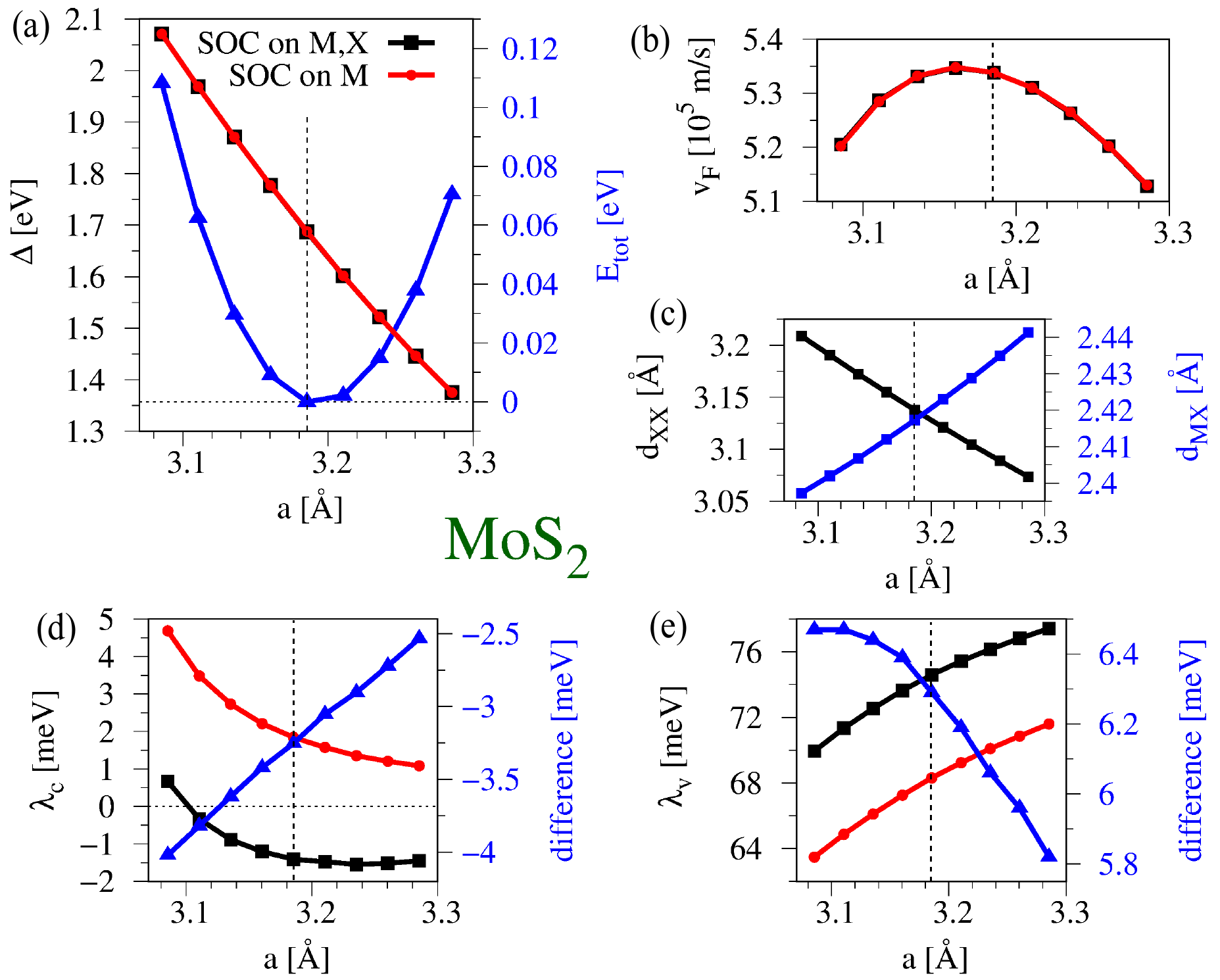}
 \caption{(Color online) Summary of the fit parameters for MoS$_2$
 as a function of the lattice constant.  
 (a) The gap parameter $\Delta$ and the total energy E$_{\textrm{tot}}$. 
 The black data (SOC on M,X) correspond to calculations where SOC is included for both atoms M and X,
 while for the red data (SOC on M), we turned off SOC on the X atoms. Dashed vertical lines indicate the equilibrium lattice constant. 
 (b) The Fermi velocity $v_{\textrm{F}}$. (c) The
 distances $d_{\textrm{XX}}$ and $d_{\textrm{MX}}$. (d,e) The SOC parameters $\lambda_{\textrm{c}}$  
 and $\lambda_{\textrm{v}}$. The difference (blue curve) is between the black and the red cure. 
 }\label{Fig:param_monolayers}
\end{figure*}

The calculated band structure of MoS$_2$ including SOC is 
shown in Fig. \ref{Fig:bands_TMDC_fit}
as a representative example for all considered TMDCs. 
In agreement with previous calculations \cite{Kormanyos2014:PRX, Liu2015:CSR, Kosmider2013:PRB, Kormanyos2014:2DM}, 
we observe the spin valley coupling at K and K' point. 
We are able to fit the Hamiltonian, $\mathcal{H}_{0}+\mathcal{H}_{\Delta}+\mathcal{H}_{\textrm{soc}}$, 
to the low energy bands of the TMDC at K and K' valley and
obtain a very good agreement with the calculated band structure, as can be seen in Figs. \ref{Fig:bands_TMDC_fit}(b,c).
The fit parameters for the different TMDCs are summarized in Table \ref{Tab:alat_TMDC}, considering the equilibrium lattice constants obtained from first-principles lattice relaxation.

In order to analyze the dependence on the lattice constant, i. e., biaxial strain, 
we allow the chalcogen atoms to relax in their $z$ position, 
for every considered lattice constant. Therefore, we do not change the
symmetry, but naturally the distances $d_{\textrm{XX}}$ and $d_{\textrm{MX}}$ will change, 
as we apply biaxial strain. 
We then calculate the low energy band structure around the K and K' valleys
and fit the model Hamiltonian $\mathcal{H}_{0}+\mathcal{H}_{\Delta}+\mathcal{H}_{\textrm{soc}}$, 
for a series of lattice constants. 
Due to time-reversal symmetry, it is enough to fit the Hamiltonian around the K point, 
taking into account the spin expectation values of the bands in order to find the 
correct signs of $\lambda_{\textrm{c}}$ and $\lambda_{\textrm{v}}$.  
The three parameters $\Delta$, $\lambda_{\textrm{c}}$, and $\lambda_{\textrm{v}}$ are fitted at the K point, where we have four DFT-energies and three energy differences. 
The remaining parameter $v_{\textrm{F}}$ is fitted around the K point, 
to capture the curvature of the bands. The fitted parameters are thus free from correlations.
In Fig. \ref{Fig:param_monolayers} we show the fit parameters obtained for MoS$_2$ as function of the lattice constant.  
We find that the total energy E$_{\textrm{tot}}$ is minimized 
for the DFT predicted lattice constant \cite{Kormanyos2014:2DM}, 
which slightly deviate from the experimentally determined one for a bulk TMDC, 
also listed in Table \ref{Tab:alat_TMDC}.

As we vary the lattice constant from smaller to larger values the distance between two 
chalcogen atoms $d_{\textrm{XX}}$ is getting smaller, while the distance between the transiton metal 
atom and the chalcogen atom $d_{\textrm{MX}}$ is getting larger, see Fig. \ref{Fig:param_monolayers}(c). 
The parameter $\Delta$, describing the orbital gap at K and K' valley, decreases as we increase the 
lattice constant in agreement with literature \cite{Chang2013:PRB, Wang2014:AdP, Frisenda2017:2DMA, Johari2012:ACS, Muoi2019:CP, Ahn2017:NC}. 
Keeping the orbital decomposed band structure (Fig. \ref{Fig:bands_TMDC_character}) in mind, the lattice constant influences all atomic distances, the overlap of $p$ and $d$-orbitals, and matrix elements in a tight-binding model perspective \cite{Cappelluti2013:PRB, Liu2013:PRB}.
Consequently, the energy of a given band at a certain $k$-point 
changes with the atomic distances. 
For example, the CB (VB) edge at the K ($\Gamma$) point is formed by $d_{z^2}$-orbitals and
shifts down (up) in energy with increasing lattice constant, see animations in the Supplemental Material \footnotemark[1]. 
 
Note that for MoS$_2$ and strains of about $-1$\% (+1\%), when we have a smaller (larger) lattice constant, the band gap becomes indirect \cite{Muoi2019:CP, Johari2012:ACS, Wang2014:AdP} 
and is at K $\rightarrow$ Q ($\Gamma \rightarrow$ K), where Q is the 
CB side valley along the K-$\Gamma$ line, see Fig. \ref{Fig:bands_TMDC_character}.
For the other TMDCs, the situation is similar, but for different strain amplitudes. 
Tuning the gap with uniaxial or biaxial strain consequently modifies the optical properties, such as the photolominescence spectrum, exciton-phonon coupling and circular dichroism 
\cite{Niehues2018:NL, Feierabend2017:PRB, Aslan2018:PRB, Aas2018:OE}. 
It has also been shown that strain applied to MoS$_2$-based photodetectors can control the response time of the devices \cite{Gant2019:MatToday}. We address the effects of strain in the direct-indirect optical transitions in Sec.~\ref{sec:Transitions} and the role of excitonic 
effects in the direct gap regime in Sec.~\ref{sec:Excitons}.

\begin{table*}[htb]
\caption{Fit parameters of the model Hamiltonian Eq. (\ref{Eq:Hamiltonian}) for all four TMDCs and different values of biaxial strain. The lattice parameter $a$ is given in \AA, $\Delta$ is given in 
eV, $v_{\textrm{F}}$ is given in $10^{5}$~m/s and $\lambda_{\textrm{c}}$, $\lambda_{\textrm{v}}$ are given in meV.}
\label{Tab:all_params}
\begin{ruledtabular}
\begin{tabular}{ccccccccccc}
\multicolumn{5}{c}{MoS$_2$} & & \multicolumn{5}{c}{MoSe$_2$} \\
\cline{1-5} \cline{7-11}
$a$ & $\Delta$ & $v_{\textrm{F}}$ & $\lambda_{\textrm{c}}$ & $\lambda_{\textrm{v}}$ & & $a$ & $\Delta$ & $v_{\textrm{F}}$ & $\lambda_{\textrm{c}}$ & $\lambda_{\textrm{v}}$ \\
\cline{1-5} \cline{7-11}
3.0854 & 2.071 & 5.205 &  0.666 & 69.95  & &  3.219 & 1.779 & 4.429 & -9.120 & 89.10 \\
3.1104 & 1.969 & 5.287 & -0.336 & 71.34  & &  3.244 & 1.696 & 4.507 & -10.22 & 90.39 \\
3.1354 & 1.871 & 5.331 & -0.887 & 72.54  & &  3.269 & 1.615 & 4.560 & -10.59 & 91.54 \\
3.1604 & 1.777 & 5.347 & -1.207 & 73.64  & &  3.294 & 1.536 & 4.587 & -10.64 & 92.50 \\
3.1854 & 1.687 & 5.338 & -1.410 & 74.60  & &  3.319 & 1.461 & 4.597 & -10.45 & 93.25 \\
3.2104 & 1.602 & 5.310 & -1.479 & 75.43  & &  3.344 & 1.389 & 4.589 & -10.15 & 93.95 \\
3.2354 & 1.522 & 5.263 & -1.550 & 76.16  & &  3.369 & 1.320 & 4.564 & -9.760 & 94.42 \\
3.2604 & 1.446 & 5.202 & -1.518 & 76.82  & &  3.394 & 1.255 & 4.526 & -9.309 & 94.78 \\
3.2854 & 1.375 & 5.128 & -1.450 & 77.42  & &  3.419 & 1.194 & 4.480 & -8.780 & 94.98 \\
\\
\multicolumn{5}{c}{WS$_2$} & & \multicolumn{5}{c}{WSe$_2$} \\
\cline{1-5} \cline{7-11}
$a$ & $\Delta$ & $v_{\textrm{F}}$ & $\lambda_{\textrm{c}}$ & $\lambda_{\textrm{v}}$ & & $a$ & $\Delta$ & $v_{\textrm{F}}$ & $\lambda_{\textrm{c}}$ & $\lambda_{\textrm{v}}$ \\
\cline{1-5} \cline{7-11}
3.080 & 2.274 & 6.752 & 43.08 & 191.46  & &  3.219 & 1.917 & 5.964 & 51.18 & 212.16 \\
3.105 & 2.153 & 6.815 & 31.80 & 197.68  & &  3.244 & 1.816 & 6.011 & 38.40 & 218.12 \\
3.130 & 2.035 & 6.820 & 24.40 & 203.40  & &  3.269 & 1.716 & 6.019 & 29.87 & 223.60 \\
3.155 & 1.921 & 6.795 & 19.35 & 208.64  & &  3.294 & 1.619 & 5.995 & 24.00 & 228.56 \\
3.180 & 1.812 & 6.735 & 15.72 & 213.46  & &  3.319 & 1.525 & 5.948 & 19.86 & 233.07 \\
3.205 & 1.710 & 6.655 & 13.07 & 217.83  & &  3.344 & 1.437 & 5.881 & 16.85 & 237.10 \\
3.230 & 1.614 & 6.542 & 11.11 & 221.81  & &  3.369 & 1.353 & 5.798 & 14.63 & 240.72 \\
3.255 & 1.526 & 6.437 &  9.62 & 225.37  & &  3.394 & 1.276 & 5.702 & 13.02 & 243.86 \\
3.280 & 1.443 & 6.311 &  8.46 & 228.57  & &  3.419 & 1.203 & 5.596 & 11.81 & 246.58 \\
\end{tabular}
\end{ruledtabular}
\end{table*}

The Fermi velocity, $v_{\textrm{F}}$, reflecting the effective mass, does not change drastically 
as we vary the lattice constant, but still we see some characteristic nonlinear behavior, see 
Fig. \ref{Fig:param_monolayers}(b). The reason is that $v_{\textrm{F}} \propto a t$ \cite{Xiao2012:PRL}, given by the effective hopping integral $t$ between 
$d_{z^2}$ and $d_{xy+x^2-y^2}$ orbitals, mediated by chalcogen $p$ orbitals,  
is influenced by atomic distances $d_{\textrm{XX}}$ and $d_{\textrm{MX}}$. 
The most interesting are the SOC parameters $\lambda_{\textrm{c}}$ and 
$\lambda_{\textrm{v}}$, see Figs. \ref{Fig:param_monolayers}(d,e). 
Because we have two different atomic species in the unit cell, we consider the influence of the individual 
atoms, M and X, on the SOC parameters, which represent the spin-splittings of the CB and VB. 
For that we calculate the band structure once with SOC on both atom species 
and once artificially turning off SOC on the chalcogen atom, 
by using a non-relativistic pseudopotential for it. 
This allows us to resolve the contributions from the M and X atom to the SOC parameters individually.
The difference (blue curve) reflects the contribution from the chalcogen 
atoms to the splittings, see Figs. \ref{Fig:param_monolayers}(d,e). 

We find that the parameter $\lambda_{\textrm{c}}$ decreases, while
the parameter $\lambda_{\textrm{v}}$ increases with increasing lattice constant. 
Both parameters depend in a nonlinear fashion on the biaxial strain. 
At a certain lattice constant, the CB splitting in MoS$_2$ can even make a
transition through zero, reordering the two spin-split bands in Fig. \ref{Fig:bands_TMDC_fit}(b). 
In addition, the differences (blue curve) in $\lambda_{\textrm{c}}$ and 
$\lambda_{\textrm{v}}$ decrease in magnitude, as we increase the lattice constant. 
This we can understand from the fact, that the spin-splittings 
of the CB and VB result from an interplay of the
atomic SOC values of the transition-metal and chalcogen atoms, 
as derived from perturbation theory in Ref. \cite{Kosmider2013:PRB}. 
We confirm that the chalcogen atom has a negative contribution to the
CB splitting and a positive contribution to the VB splitting, while the 
transition metal atom gives positive contributions to both
splittings, in agreement with Ref. \cite{Kosmider2013:PRB}. 
In Fig. \ref{Fig:param_monolayers} we explicitly show, 
how the spin splittings depend on the lattice constant
and how the different atom types contribute to it for the case of MoS$_2$.
What is still missing so far, is a microscopic orbital-based description of
how the spin splittings depend on the lattice constant and respective distances, 
as for example derived for graphene \cite{Konschuh2010:PRB}.  

In a fashion similar to Fig. \ref{Fig:param_monolayers} 
we calculate the same dependence on the
lattice constant for other TMDCs, see Supplemental Material \footnotemark[1].  
For all of them, we can observe similar characteristic trends of the parameters, 
varying as function of the lattice constant. 
The fitted parameters as a function of the lattice constant are 
summarized in Table \ref{Tab:all_params} for all TMDCs.
An interesting observation is that the CB SOC parameter $\lambda_{\textrm{c}}$ 
for Mo-based systems is opposite in sign compared 
to W-based materials, as already pointed out 
in earlier works \cite{Kormanyos2014:2DM,Kormanyos2014:PRX, Kosmider2013:PRB}.

In the Supplemental Material \footnotemark[1] we provide animations that explicitly show the evolution
of the TMDC band structures as function of biaxial strain. Additionally, we compare the results for
all TMDCs obtained from two different exchange correlation functionals, namely PBE \cite{Perdew1996:PRL}
and PBEsol \cite{Perdew2008:PRL}. In the case of PBEsol, which improves equilibrium properties, the
total energy is minimized for the experimental lattice constant. However, the overall magnitudes and
trends of the parameters as function of the lattice constant, are barely different. 
We conclude that the PBEsol functional should hardly influence the following results on exciton energy
levels and gauge factors, and results can be compared to experiment, when regarding them relative to 0\% strain (equilibrium lattice constant). 

\section{Strain tunable optical transitions}

\subsection{Direct and indirect band gap regimes}
\label{sec:Transitions}

\begin{figure}[htb]
\includegraphics[width=.99\columnwidth]{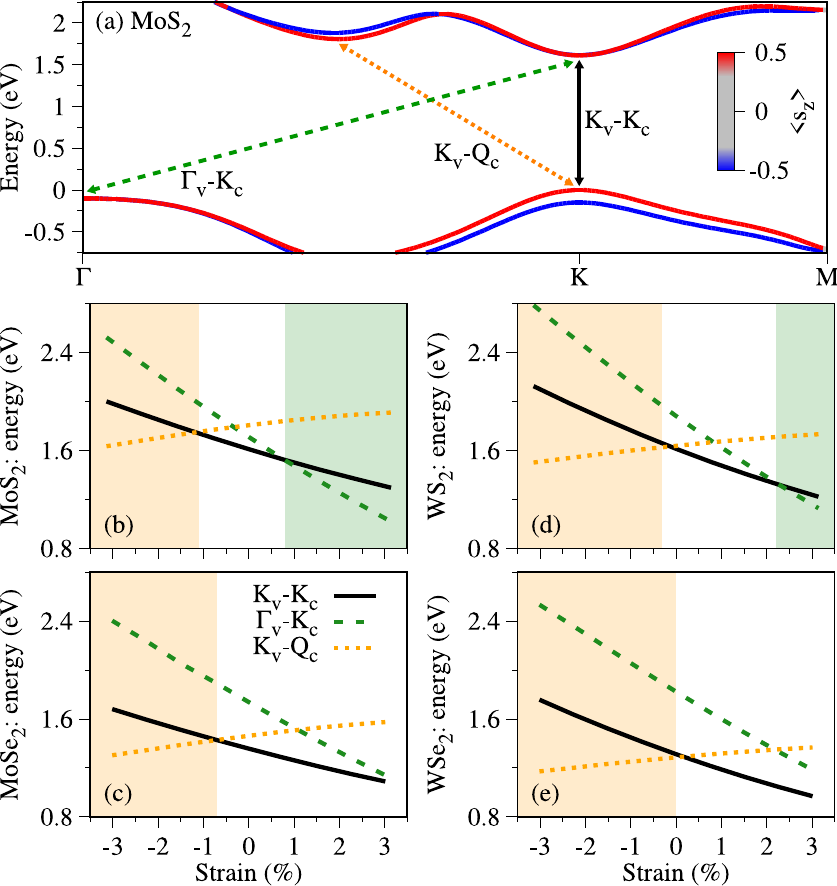}
\caption{(Color online) (a) Band structure of MoS$_2$ at zero strain highlighting 
the important optical transitions mostly affected by strain. The transitions are 
identified by the reciprocal space point (K, Q, $\Gamma$) and by the energy band 
(subindices v and c stand for the valence and conduction bands, respectively). 
Evolution of the transition energies depicted in (a) as a function of 
strain for (b) MoS$_2$, (c) MoSe$_2$, (d) WS$_2$ and (e) WSe$_2$. The shaded 
regions indicate indirect band gap regimes (K-Q for negative strain and 
$\Gamma$-K for positive strain).}
\label{Fig:transitions}
\end{figure}

In the previous section we analyzed the strain effects in the band structure 
of MoS$_2$ and found that different strain regimes induce a direct to indirect band 
gap transition. This feature is also present in the other TMDCs we investigated 
(see Supplemental Material \footnotemark[1]). In order to obtain a deeper insight into 
this direct to indirect band gap switching, in this section we discuss the strain 
dependence of the single-particle optical transitions for all the TMDCs. We focus on 
the mostly affected optical transitions, depicted in Fig.~\ref{Fig:transitions}(a) 
for MoS$_2$. The evolution of these transitions with respect to applied strain is 
shown in  Fig.~\ref{Fig:transitions}(b-e) for MoS$_2$, MoSe$_2$, WS$_2$ and WSe$_2$, 
respectively. The overall trend is similar for all TMDCs: negative strain induces 
indirect band gap for the K$_\text{v}$-Q$_\text{c}$ transition while positive 
strain values cause the $\Gamma_\text{v}$-K$_\text{c}$ transition to have the smallest 
energy. For MoSe$_2$ and WSe$_2$ the amount of positive strain required to reach 
the $\Gamma_\text{v}$-K$_\text{c}$ indirect band gap regime would be larger than 
the region we investigated here. Additionally, K$_\text{v}$-Q$_\text{c}$ transitions 
show a positive slope while K$_\text{v}$-K$_\text{c}$ and $\Gamma_\text{v}$-K$_\text{c}$ 
show a negative slope. Although a proper comparison to uniaxial strain results may 
seem unfair due to the different lattice symmetries, it is still worth mentioning 
that $\Gamma_\text{v}$-K$_\text{c}$ transitions have a steeper dependence than 
the K$_\text{v}$-K$_\text{c}$ transitions, as observed experimentally for 
MoS$_2$ \cite{Conley2013:NL} and WS$_2$ \cite{Blundo2019:arXiv}, for instance. 
Furthermore, theoretical studies based on first-principles calculations have 
shown such dependencies not only due to uniaxial strain but also in the biaxial 
strain case \cite{Peelaers2012:PRB,Wang2014:AdP}.

One important figure of merit to analyze the strain dependence is the so called 
gauge factor of the transition energies, i. e., the rate of energy shift due to 
the applied strain, typically given in meV/\%. In Table~\ref{Tab:transitions}, we 
quantify the gauge factors for the different transition energies shown in 
Figs.~\ref{Fig:transitions}(b-e). Although these energy transitions do not behave 
completely linear under strain, we assumed for simplicity a linear behavior 
throughout the whole strain range we considered. We found that the strength of 
gauge factors for the indirect $\Gamma_\textrm{v}$-K$_\textrm{c}$ transitions is 
nearly twice as large as the direct K$_\textrm{v}$-K$_\textrm{c}$ transitions. 
On the other hand, the strength of the gauge factors of the indirect 
K$_\textrm{v}$-Q$_\textrm{c}$ transitions are nearly 2 (4) times smaller 
than the direct K$_\textrm{v}$-K$_\textrm{c}$ transitions for Mo(W)-based TMDCs. 
Such large differences in the gauge factors provide important information to 
identify the evolution of the optical spectra under applied strain.

\begin{table}[htb]
\caption{Gauge factors (in meV/\%) for the single-particle transitions presented in Fig.~\ref{Fig:transitions}, extracted by linear extrapolation within the whole considered strain region.} 
\label{Tab:transitions}
\begin{ruledtabular}
\begin{tabular}{lcccc}
 & MoS$_2$ & MoSe$_2$ & WS$_2$ & WSe$_2$ \\
\colrule
K$_\textrm{v}$-K$_\textrm{c}$      & -112.3 &  -98.2 & -133.5 & -118.8 \\
$\Gamma_\textrm{v}$-K$_\textrm{c}$ & -239.9 & -210.2 & -254.0 & -213.4 \\
K$_\textrm{v}$-Q$_\textrm{c}$      &   44.3 &   45.7 &   37.0 &   32.8
\end{tabular}
\end{ruledtabular}
\end{table}

\subsection{Excitonic effects in the direct band gap regime}
\label{sec:Excitons}

\begin{figure}[htb]
\includegraphics[width=.99\columnwidth]{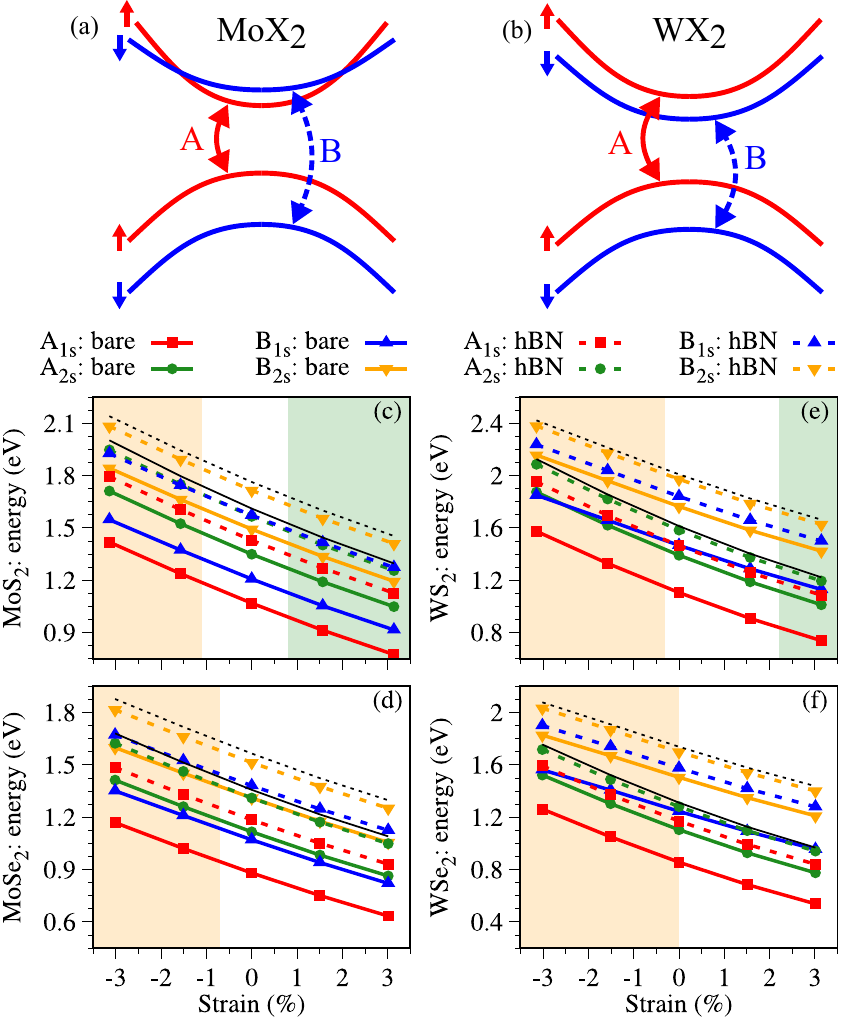}
\caption{(Color online) Sketch of the energy bands that contribute to the formation 
of A and B excitons in (a) Mo-based and (b) W-based TMDCs. Evolution of the total 
exciton energy as function of the biaxial strain for (c) MoS$_2$, (d) MoSe$_2$, 
(e) WS$_2$ and (f) WSe$_2$. The thin solid (dashed) lines are the single-particle 
energies for the A (B) optical transitions. The shaded regions indicate indirect 
band gap regimes (K-Q for negative strain and $\Gamma$-K for positive strain).}
\label{Fig:exciton_total}
\end{figure}

For moderate applied strain the direct band gap at the K point remains the fundamental 
transition energy. In this section we investigate the role of excitonic effects to 
such direct transitions under the applied biaxial strain. In a simple picture, an 
exciton is a quasi-particle created due to the electrostatic Coulomb interaction 
between electrons and holes \cite{Chuang1995book,Haug2009book}. Because of the weak 
screening of 2D materials, excitons have large binding energies and, therefore, 
excitonic effects dominate the optical
spectra \cite{Mak2010:PRL,Chernikov2014:PRL,Qiu2013:PRL,Wang2018:RMP}. Starting 
from the effective Hamiltonian given in Eq.~\eqref{Eq:Hamiltonian} and 
fitted parameters given in Table~\ref{Tab:all_params}, we compute the excitonic 
spectra of the strained monolayer TMDCs for different bright excitonic states 
(the s-like excitons) that can be directly probed in experiments. We use 
the effective Bethe-Salpeter equation (BSE) 
\cite{RohlfingPRB:2000,Scharf2017:PRL,Scharf2019:JPCM,Tedeschi2019:PRB,FariaJunior2019:PRB,Zollner2019:PRB} 
with the electron-hole interaction mediated by the Rytova-Keldysh 
potential \cite{Rytova1967coulomb,Keldysh1979coulomb,Cudazzo2011:PRB,Berkelbach2013:PRB}. 
The screening lengths of the TMDCs are taken from the study of 
Berkelbach \textit{et al.} \cite{Berkelbach2013:PRB}.
The BSE is solved on a 2D $k$-grid from -0.5 to 0.5 
$\textrm{\AA}^{-1}$ in the $k_x$ and $k_y$ directions with total discretization 
of $101 \times 101$ points (leading to a spacing of $\Delta k = 10^{-2} \; \textrm{\AA}^{-1}$). 
To improve convergence, the Coulomb potential is averaged around each $k$-point in a 
square region of $-\Delta k /2$ to $\Delta k /2$ discretized with $101 \times 101$ 
points \cite{Scharf2017:PRL,Zollner2019:PRB}.

We focus on two different exciton types: the so-called A and B excitons. 
In Mo(W)-based TMDCs, the A excitons are formed by the first VB 
and first (second) CB while B excitons are formed by the 
second VB and second (first) CB, sketched in Figs.~\ref{Fig:exciton_total}(a-b).
In Figs.~\ref{Fig:exciton_total}(c-f) we show the behavior of the total energy of A and 
B excitons as a function of the applied biaxial strain in two different dielectric environments: 
bare (effective dielectric constant of $\varepsilon = 1.0$) and hexagonal boron nitride (hBN) encapsulated TMDCs 
(effective dielectric constant of $\varepsilon = 4.5$ \cite{Stier2018:PRL}). The subindices 1s 
and 2s indicate the first and second s-like exciton states, respectively. Despite 
the nonlinear behavior of $\lambda_{\textrm{c}}$, $\lambda_{\textrm{v}}$ and $v_{\textrm{F}}$ 
seen in Fig.~\ref{Fig:param_monolayers}, the A excitons evolve in quite a linear 
fashion with the same qualitative behavior for all TMDCs. On the other hand, 
the B excitons show a different behavior for Mo and W-based TMDCs as function 
of strain. For the bare case, in Mo-based TMDCs the B exciton would be the second visible 
absorption peak while in W-based TMDCs additional peaks of the A excitons would be visible 
at energies lower than the peaks of the B excitons. Once we change the dielectric environment 
from bare to hBN-encapsulated, the ordering of the excitonic peaks changes in MoS$_\text{2}$ 
and MoSe$_\text{2}$; that is, the B exciton is no longer the second visible peak. Nevertheless, 
the same qualitative behavior as function of the biaxial strain holds, as discussed 
for the bare TMDCs case.

\begin{table}[htb]
\caption{Gauge factors (in meV/\%) for single-particle transitions and exciton levels, extracted by linear extrapolation within the $-1.5$\% to $1.5$\% strain range.
The single-particle energies for the A and B optical transitions are presented as E$_\text{A}$ and E$_\text{B}$.}
\label{Tab:exciton_total}
\begin{ruledtabular}
\begin{tabular}{lcccc}
 & MoS$_2$ & MoSe$_2$ & WS$_2$ & WSe$_2$ \\
\colrule
This work & & & & \\
E$_\text{A}$         &  -112.5  &   -98.6  &  -144.0  &  -131.2 \\
E$_\text{B}$         &  -109.8  &   -97.2  &  -123.8  &  -109.7 \\
A$_\text{1s}$: bare  &  -103.3  &   -89.6  &  -134.1  &  -121.1 \\
A$_\text{2s}$: bare  &  -106.2  &   -92.0  &  -137.8  &  -124.7 \\
B$_\text{1s}$: bare  &  -101.7  &   -89.5  &  -118.1  &  -104.4 \\
B$_\text{2s}$: bare  &  -104.1  &   -91.5  &  -120.2  &  -106.2 \\
A$_\text{1s}$: hBN   &  -106.9  &   -92.7  &  -138.6  &  -125.5 \\
A$_\text{2s}$: hBN   &  -110.4  &   -96.1  &  -142.3  &  -129.3 \\
B$_\text{1s}$: hBN   &  -104.8  &   -92.0  &  -120.6  &  -106.5 \\
B$_\text{2s}$: hBN   &  -107.9  &   -95.0  &  -122.7  &  -108.6 \\
\colrule
GW-BSE\cite{Frisenda2017:2DMA} & & & & \\
E$_\text{A}$        &   -134  & -115 &  -156 &  -141 \\
A$_\text{1s}$: bare &   -110  &  -90 &  -151 &  -134 \\
B$_\text{1s}$: bare &   -107  &  -89 &  -130 &  -111 \\
\end{tabular}
\end{ruledtabular}
\end{table}

In Table~\ref{Tab:exciton_total}, we present the gauge factors 
for the exciton peaks, i. e., the total energy given in Fig.~\ref{Fig:exciton_total}(c-f), 
extracted as a linear fit in the $-1.5$\% to $1.5$\% strain range. As a general trend, the strength 
of the gauge factors follow the order MoSe$_\text{2}$ $<$ MoS$_\text{2}$ $<$ WSe$_\text{2}$ $<$ 
WS$_\text{2}$, and the effect of changing the dielectric surroundings modifies only 2 -- 4 meV/\%, 
which can be at the scale of experimental uncertainty. Although we have not taken into account 
corrections to the band gap, our calculated exciton behaviors are in good agreement with GW-BSE 
\textit{ab-initio} calculations from Frisenda \textit{et al.}  \cite{Frisenda2017:2DMA}, also shown in 
Table~\ref{Tab:exciton_total} for comparison to our results. From the experimental perspective, 
the amount of studies on biaxial strain is still very scarce and mainly limited to MoS$_\text{2}$. 
For the available gauge factors in MoS$_\text{2}$, Plechinger \textit{et al.}  \cite{Plechinger2015:2DMat} 
found -105 meV/\% for the A exciton, Lloyd \textit{et al.}  \cite{Lloyd2016:NL} -99 $\pm$ 6 meV/\% for both 
A and B excitons and Gant \textit{et al.}  \cite{Gant2019:MatToday} a value of -94 meV/\% for the A exciton. 
Furthermore, the study of Frisenda \textit{et al.}  \cite{Frisenda2017:2DMA} also determined experimentally 
the gauge factor of MoSe$_\text{2}$, MoS$_\text{2}$, WSe$_\text{2}$ and WS$_\text{2}$ but the 
values are smaller than the theoretical results, most likely because the strain present 
in the substrate is not fully transferred to the TMDC and the calibration is not a straightforward 
task, as already discussed by the authors \cite{Frisenda2017:2DMA}.

\begin{figure}[htb]
\includegraphics[width=.99\columnwidth]{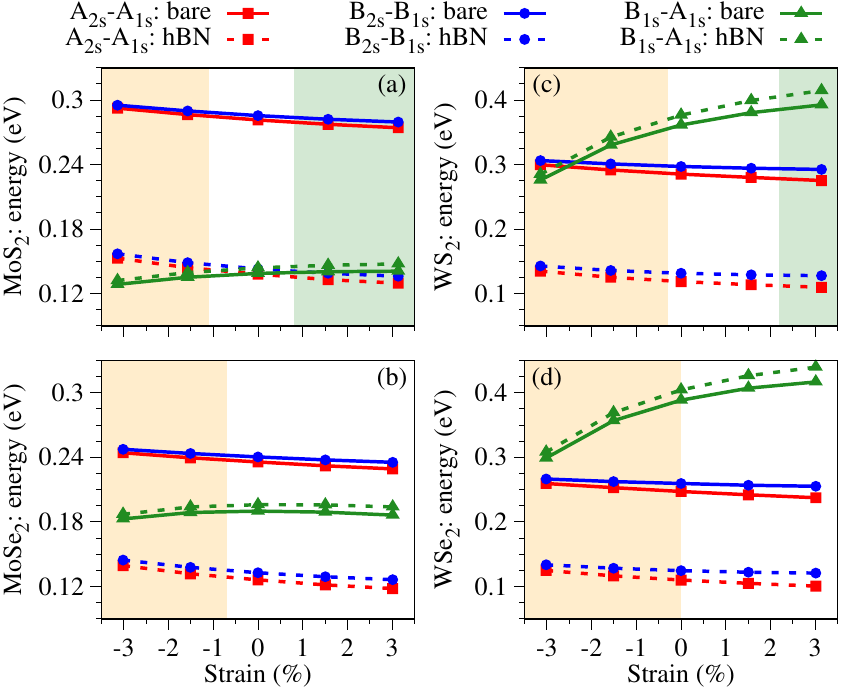}
\caption{(Color online) Evolution of the energy difference between distinct excitonic 
levels as function of the biaxial strain for (a) MoS$_2$, (b) MoSe$_2$, 
(c) WS$_2$ and (d) WSe$_2$. The shaded regions have the same meaning as in 
Fig.~\ref{Fig:exciton_total}.}
\label{Fig:exciton_ediff}
\end{figure}

\begin{table}[htb]
\caption{Extracted gauge factors (in meV/\%) for energy difference of single-particle 
transitions and excitonic levels.}
\label{Tab:exciton_ediff}
\begin{ruledtabular}
\begin{tabular}{lcccc}
 & MoS$_2$ & MoSe$_2$ & WS$_2$ & WSe$_2$ \\
\colrule
E$_\text{B}$-E$_\text{A}$          &     2.7  &     1.4  &    20.2  &    21.5 \\
A$_\text{2s}$-A$_\text{1s}$: bare  &    -2.9  &    -2.5  &    -3.7  &    -3.6 \\
A$_\text{2s}$-A$_\text{1s}$: hBN   &    -3.5  &    -3.4  &    -3.7  &    -3.9 \\
B$_\text{2s}$-B$_\text{1s}$: bare  &    -2.5  &    -2.0  &    -2.1  &    -1.8 \\
B$_\text{2s}$-B$_\text{1s}$: hBN   &    -3.1  &    -2.9  &    -2.1  &    -2.1 \\
B$_\text{1s}$-A$_\text{1s}$: bare  &     1.6  &     0.1  &    16.0  &    16.7 \\
B$_\text{1s}$-A$_\text{1s}$: hBN   &     2.1  &     0.7  &    18.0  &    18.9 \\
\end{tabular}
\end{ruledtabular}
\end{table}

Besides the total exciton energies, it is also helpful to look at how the energy 
separation of different excitonic levels change under the applied strain. These behaviors 
are summarized in Fig.~\ref{Fig:exciton_ediff} for all TMDCs considered here and 
the corresponding gauge factors are presented in Table~\ref{Tab:exciton_ediff}. 
Although the change in the dielectric environment has a minor effect on the gauge factors 
(2 meV/\% or less), it drastically changes the total energy difference by hundreds of meV for 
the  A$_\text{2s}$-A$_\text{1s}$ and B$_\text{2s}$-B$_\text{1s}$ exciton separation (compare
solid and dashed lines with squares and circles in Fig.~\ref{Fig:exciton_ediff}). 
On the other hand the energy separation of B$_\text{1s}$-A$_\text{1s}$ excitons is affected 
by only a few or tens of meV (compare solid and dashed lines with triangles 
in Fig.~\ref{Fig:exciton_ediff}). Furthermore, the gauge factor of 
B$_\text{1s}$-A$_\text{1s}$ energy difference for W-based compounds is one order of magnitude 
larger than that of the Mo-based compounds, reflecting the larger increase of $\lambda_\textrm{v}$ 
(see for instance Fig.~\ref{Fig:param_monolayers}). We point out that for WSe$_2$ our calculations 
reveal the same qualitative trends as in recent experiments with uniaxial strain by 
Aslan \textit{et al.} \cite{Aslan2018:PRB}, in which they found a gauge factor of -6$\pm$1 meV/\% for 
the A$_\text{2s}$-A$_\text{1s}$ exciton separation and 10 meV/\% for 
B$_\text{1s}$-A$_\text{1s}$ exciton separation.

\section{Summary}
\label{sec:Summary}

We have shown that applying biaxial strain to monolayer TMDCs induces drastic changes 
in their orbital, spin-orbit and, consequently, optical properties. Furthermore, 
we showed on a quantitative level, how the spin-orbit band 
splittings in a TMDC depend on biaxial strain and on the SOC contributions from 
the individual atoms. Additionally, by employing the Bethe-Salpeter equation 
combined with a minimal tight-binding Hamiltonian fitted to the {\it ab-initio} 
band structure, we have calculated the evolution of several direct exciton peaks 
as a function of biaxial strain and for different dielectric surroundings. 
Specifically, we found that the gauge factors are slightly affected by the dielectric 
environment and  are mainly ruled by the atomic composition, with the ordering MoSe$_\text{2}$ $<$ MoS$_\text{2}$ $<$ WSe$_\text{2}$ $<$ WS$_\text{2}$. Our 
results provide valuable insights into how strain can modify the TMDC properties 
within van der Waals heterostructures, and the parameter sets we provided can be 
applied to investigate other physical phenomena.

\footnotetext[1]{See Supplemental Material at [URL will be inserted by publisher] including Refs. \cite{Perdew2008:PRL,Perdew1996:PRL}, where we show band structure animations, further fit results for WS$_2$, WSe$_2$, and MoSe$_2$ as function of the lattice constant, and a comparison between PBE and PBEsol functional for all TMDCs.}

\acknowledgments
We thank A. Polimeni for helpful discussions. 
This work was supported by DFG SPP 1666, DFG SFB 1277 (project B05), the European Unions Horizon 2020 research and innovation program under Grant No. 785219, the Alexander von Humboldt Foundation and Capes (Grant No. 99999.000420/2016-06).

\bibliography{paper}

\cleardoublepage
\includepdf[pages=1]{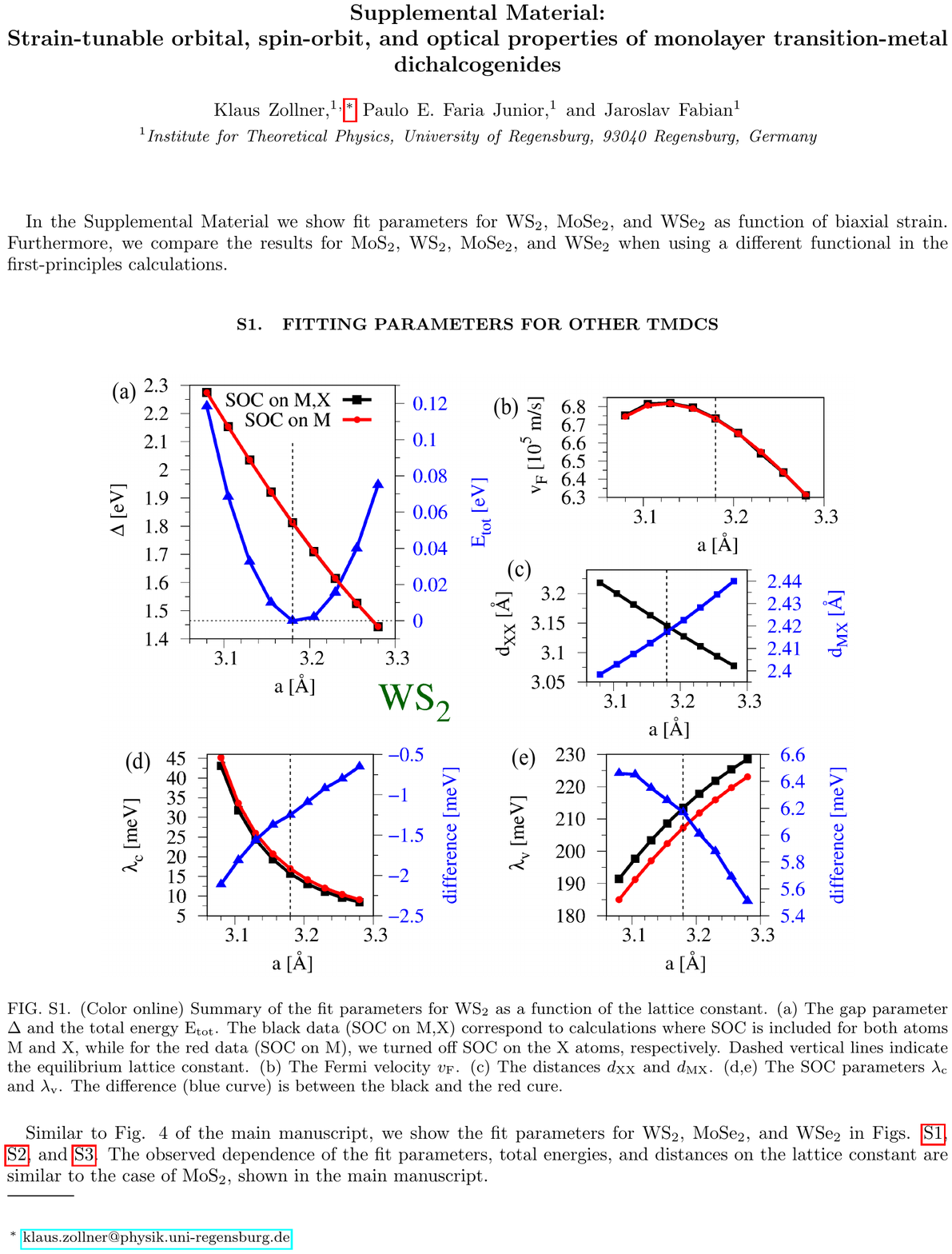}\clearpage
\includepdf[pages=2]{supplement.pdf}\clearpage
\includepdf[pages=3]{supplement.pdf}\clearpage
\includepdf[pages=4]{supplement.pdf}\clearpage
\includepdf[pages=5]{supplement.pdf}\clearpage
\includepdf[pages=6]{supplement.pdf}\clearpage
\includepdf[pages=7]{supplement.pdf}\clearpage

\end{document}